\documentclass[prb,twocolumn,showpacs,amsmath,amssymb,floatfix,superscriptaddress]{revtex4}
\usepackage{amsmath}
\usepackage{float}
\usepackage{graphicx}
\usepackage{gensymb}
\newcommand{\hef}{$^4$He}

\newcommand{\het}{$^3$He }

\newcommand{\pA}{\text{\AA}^{-2}}

\newcommand{\rholp}{{\rho_1^{\text{lp}}}}
\newcommand{\K}{\text{K}}
\newcommand{\A}{\text{\AA}}

\begin{document}
\title{Phase diagram of $^4$He adsorbed on graphite}
\author{Philippe Corboz}
\affiliation{Institut f{\"u}r theoretische Physik, ETH Z\"urich, CH-8093 Z{\"u}rich, Switzerland}
\author{Massimo Boninsegni}
\affiliation{Department of Physics, University of Alberta, Edmonton, Alberta, Canada T6G 2J1}
\author{Lode Pollet}
\affiliation{Institut f{\"u}r theoretische Physik, ETH Z\"urich, CH-8093 Z{\"u}rich, Switzerland}
\author{Matthias Troyer}
\affiliation{Institut f{\"u}r theoretische Physik, ETH Z\"urich, CH-8093 Z{\"u}rich, Switzerland}

\begin{abstract}
We present results of a theoretical study of $^4$He films adsorbed on graphite, based on the continuous space worm algorithm. In the first layer, we find a domain-wall phase and a (7/16) registered structure between the 
commensurate (1/3) and the incommensurate solid phases. For the second layer, we find only superfluid and incommensurate solid phases. The commensurate phase found  in previous simulation work is only observed if first layer particles are kept fixed; it disappears upon explicitly including their zero-point fluctuations. No evidence of any ``supersolid" phase is found.
\end{abstract}
%67.25.-k 	4He
%67.25.dp 	Films
%67.25.D- 	Superfluid phase 
%----
%67.80.bd 	Superfluidity in solid 4He, supersolid 4He
%67.80.-s 	Quantum solids
%67.80.kb 	Supersolid phases on lattices
%67.80.dm 	Films
%05.30.Jp 	Boson systems 
%02.70.Ss 	Quantum Monte Carlo methods 
 
%Manousakis PACS:
%67.70.+n,	Films (including physical adsorption) << not in use anymore
%, 67.40.Kh 	Thermodynamic properties << not in use anymore
% Massimo's paper: 67.70.+n, 05.30.Jp, 68.43.Fg (	Adsorbate structure (binding sites, geometry) ), 02.70.Ss


\maketitle
\section{Introduction}
Adsorption of helium on a graphite substrate is still the subject of many experimental and theoretical studies; although this subject is almost four decades old, it has lately been enjoying a resurgence of interest in connection with the study of a possible supersolid phase of matter. 

Due to the strong attraction to graphite, helium forms up to seven distinct layers above the substrate,\cite{Greywall93, Zimmerli92} each layer being a realization of a quasi-two dimensional system. Several types of phases result from the interplay between the interaction among helium atoms, and their interaction with the substrate, including fluid,  commensurate and incommensurate solid phases.
Crowell and Reppy raised the possibility of a supersolid phase in the second layer, from the anomalous behavior of the period shift in torsional oscillator experiments.\cite{Crowell} Supersolids exhibit simultaneously crystalline order and frictionless flow in a single homogeneous phase, and have attracted increasing interest since the observation of nonclassical moment of inertia in solid $^4$He by Kim and Chan.\cite{Kim04} It has been proposed, based on a number of fundamental arguments, that a commensurate perfect single crystal of $^4$He ought not be supersolid, 
%(cite also paper by Clark & Ceperley, Boninsegni & Prokof'ev & Svistunov
%, 
\cite{Prokofev05} a prediction supported by a number of computer simulations.\cite{Boninsegni06d,Clark06}  However, a supersolid phase exists for bosonic models on a triangular lattice \cite{Boninsegni05}
%, Heidarian05, Wessel05}, 
and one may thus speculate about the possibility of supersolids on substrates.\cite{Saunders}

For the first adsorbed helium layer, there exist  preferred adsorption sites, located above the centers of the hexagons formed by the carbon atoms on the graphite surface.
%The commensurate phase at a filling $1/3$ and the incommensurate solid phase are clearly observed  in neutron diffraction experiments \cite{Carneiro81, Lauter87, Lauter91} and heat capacity measurements \cite{Bretz73, Hering76, Greywall91}. %Ecke83
A commensurate phase at filling $1/3$, as well as an incommensurate solid phase are clearly observed  in neutron diffraction experiments,\cite{Carneiro81, Lauter87, Lauter91} heat capacity measurements, \cite{Bretz73, Hering76, Greywall91} and in numerical simulations.\cite{Pierce00} %Ecke83
The identification of the phases occurring between the two solids is still uncertain.\cite{Bruch97} Several types of domain-wall phases have been predicted,\cite{Halpin86, Greywall93} but none unambiguously observed.

The second layer is known to exhibit a gas, a superfluid and an incommensurate solid phase, as shown by heat capacity measurements \cite{Polanco78,Greywall91} and neutron diffraction experiments.\cite{ Carneiro81, Lauter87, Lauter91} At intermediate density between these two phases, Greywall and Busch conjectured a commensurate solid with a $\sqrt{7}\times \sqrt{7}$ partial registry with respect to the first layer, based on their heat capacity measurements.\cite{Greywall91,Greywall93} At the same filling, a commensurate solid phase was observed in path integral Monte Carlo (PIMC) simulations;\cite{Pierce98} in that study, however,  for computational convenience first layer particles were treated as classical, i.e., held fixed in space at their $T$=0 equilibrium position. Because of the relative weakness of the adsorption potential experienced by second layer atoms, it is plausible that the explicit inclusion of zero-point motion of first layer particles, possibly leading to a further weakening of the attraction, may qualitatively alter the picture. Moreover, no prediction has yet been made theoretically regarding the existence of a possible supersolid phase in the second layer.

In this paper, we study the low temperature phase diagram of the first and second layers of $^4$He  on graphite, by means of state-of-the-art computer simulations in which quantum zero-point motion of helium atoms in both the first and second layers is fully included.   
%
%%%%%%%%%%
%%%FIRST LAYER
%%%%%%%%%%
%%%%%%%%%%
\begin{figure}[b]
\center
  \includegraphics[width=0.48\textwidth]{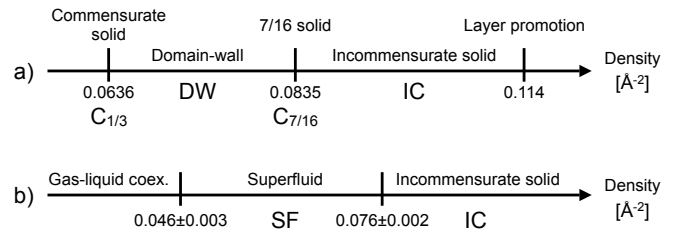}
   \caption{Schematic phase diagram of the first (a) and second (b) layers, as a function of two-dimensional layer density.}
\label{fig:firstlayer}
\end{figure}
%%%%%%%%%%%%%%%%%%%%%%%%

Our computed phase diagram is summarized in Fig.~\ref{fig:firstlayer}. In the first layer we find a striped phase, a hexagonal domain-wall phase, and a hexagonal commensurate structure (at filling 7/16) between the two crystals. In the second layer,  we do not observe a commensurate solid phase,\cite{notecommensurate} in contrast to the predictions by Greywall and Busch.\cite{Greywall91,Greywall93} Rather,  as coverage is increased the system goes through coexistence of liquid and gas phases, a (superfluid) liquid and an incommensurate crystal. 

We show that the commensurate phase observed in Ref. \onlinecite{Pierce98} is merely a  consequence of a the neglect of zero-point motion of first layer \hef\ atoms, and that such a phase disappears if this approximation is removed, i.e., if zero-point motion of first layer particles is included in the simulation. In {\it no} case is a supersolid phase of \hef\ observed, i.e., including when a commensurate second layer solid phase is ``artificially" stabilized by holding first layer atoms at fixed positions, as done in Ref. \onlinecite{Pierce98}. Not surprisingly, no supersolid phase occurs in {\it incommensurate} solid films, in analogy with what observed in three dimensions.

In the next section we briefly illustrate our model and computational methodology; we then describe our results in detail, for the first and second layer.

\section{Methodology}
Our computer simulations are based on the continuous-space worm algorithm.\cite{Boninsegni06} This methodology has proven remarkably effective in large-scale simulations of Bose systems, owing to its efficiency in sampling multi-particle exchanges, which underlie phenomena such as Bose-Einstein Condensation and superfluidity.

The microscopic model utilized here is standard. Specifically, 
we use the helium Aziz interatomic potential,\cite{Aziz79,Piercenote} as well as the anisotropic 6-12 graphite-helium potential of Carlos and Cole,\cite{Carlos79} used in essentially all previous simulation work.\cite{Whitlock98} Such a potential accounts for the corrugation of the graphite substrate, a crucial ingredient to describe the variety of registered phases in the first layer. Effects of the corrugation of the graphite substrate become negligible for successive adlayers; for simulations of the second layer we have therefore utilized the laterally averaged version of the Carlos-Cole potential (see also Ref. \onlinecite{Pierce98}).

%, and possibly second adsorbed layers.
Our model is fully three-dimensional, we use standard periodic boundary conditions, and simulate systems comprising up to 600 particles, at temperatures as low as 0.2 K, which is low enough to yield essentially ground state estimates. 
\section{Results: First Layer}
\begin{figure}[b]
\center
  \includegraphics[width=0.23\textwidth]{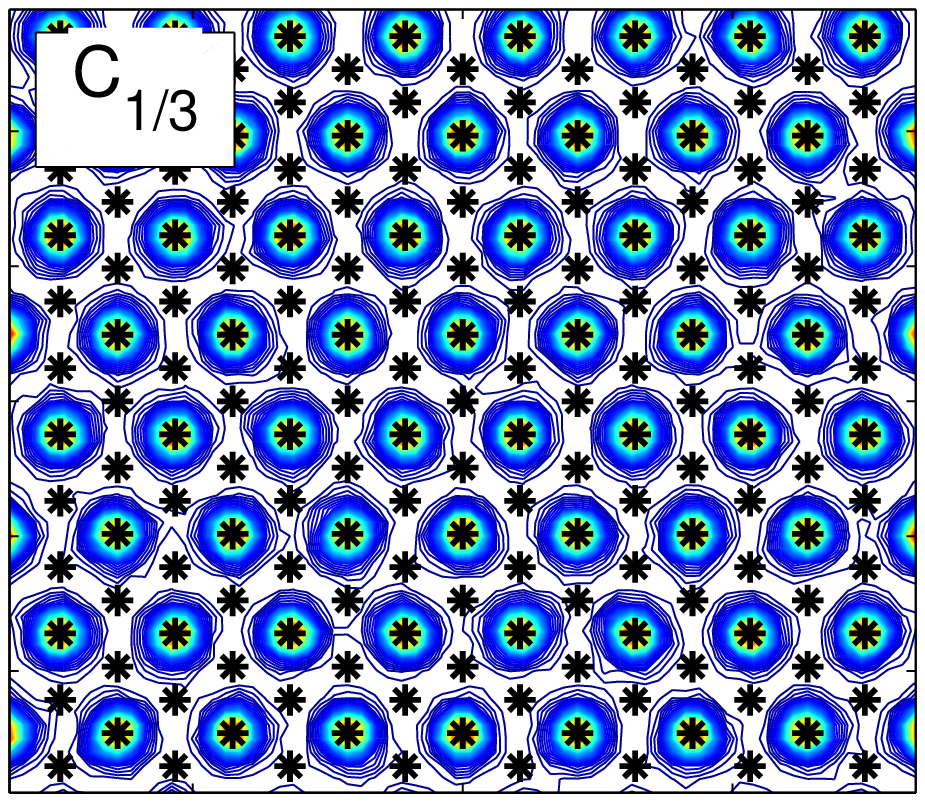}
    \includegraphics[width=0.23\textwidth]{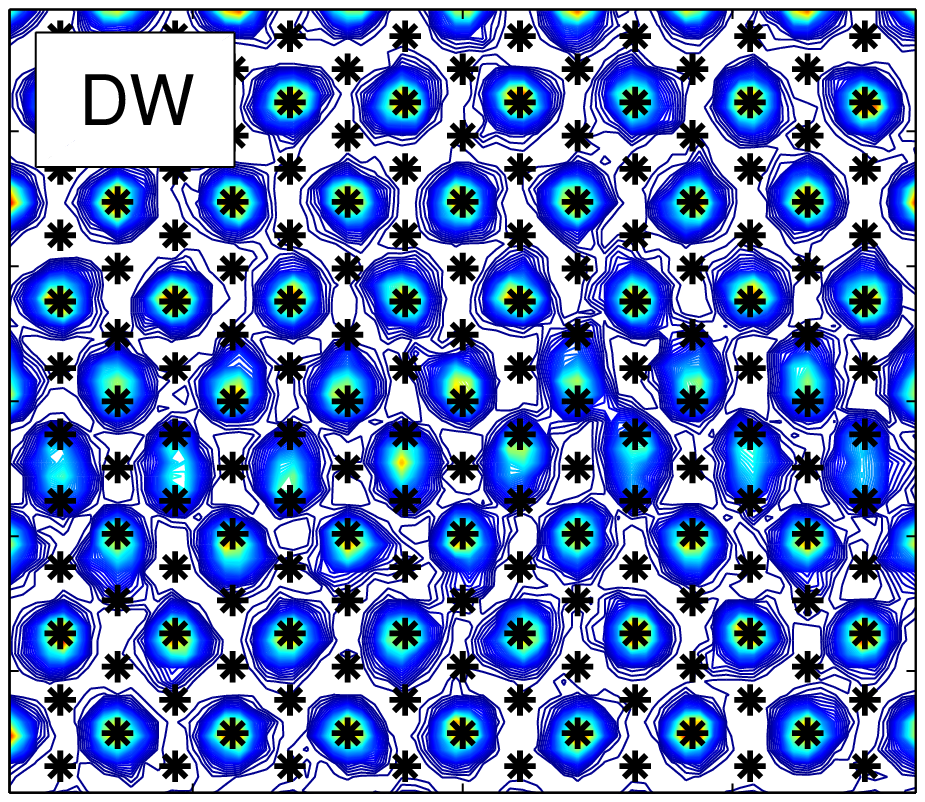}
      \includegraphics[width=0.23\textwidth]{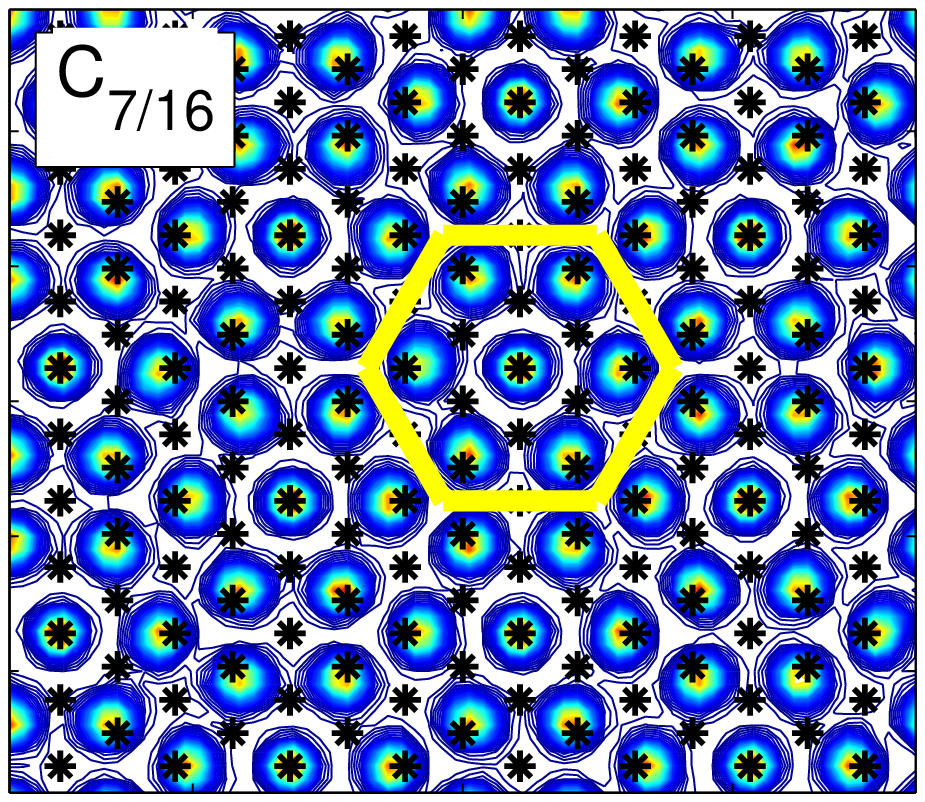}
        \includegraphics[width=0.23\textwidth]{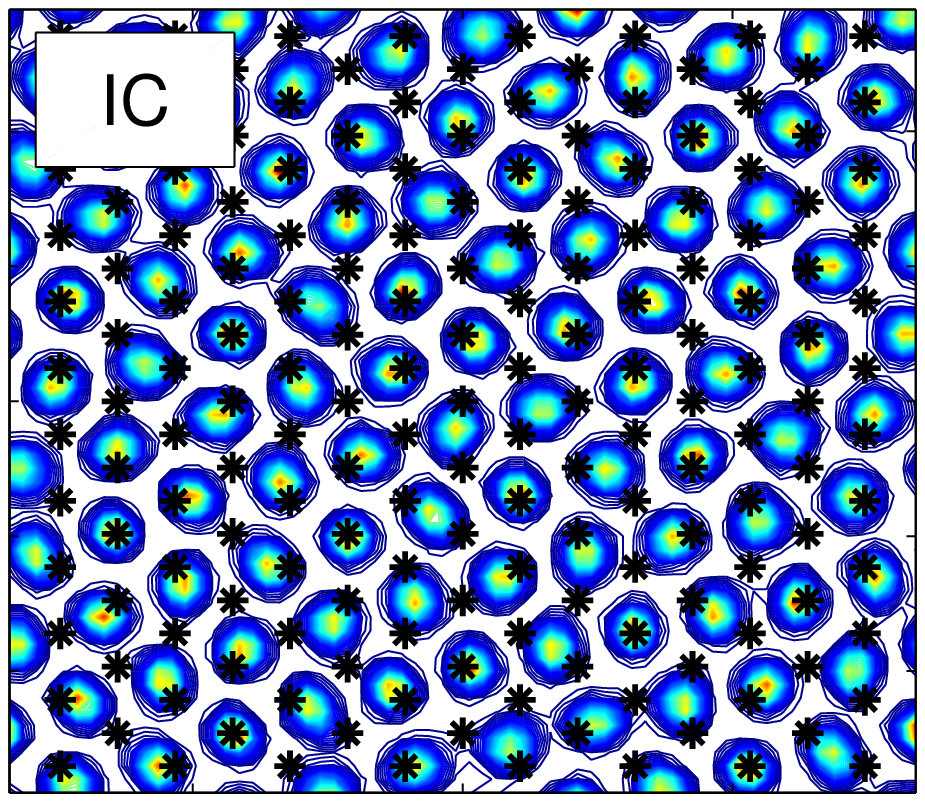}
  \caption{(Color online) Contour plots of the density distribution of the \hef atoms in the first layer (x-y-plane parallel to the substrate).  The black stars mark the minima of the underlying graphite potential (adsorption sites). The distance between two adsorption sites is $2.46\A$. The snapshots are taken at different densities in the first layer (see Fig. \ref{fig:firstlayer}): $C_{1/3}$: commensurate $1/3$ solid, $DW$:  domain-wall phase, $C_{7/16}$: commensurate $7/16$ solid and $IC$: incommensurate solid.}
\label{fig:fls}
\end{figure}

In the first layer  we confirm the existence of a commensurate $C_{1/3}$ solid at coverage $\rho_{1/3}=0.0636\pA$ (Fig. \ref{fig:fls}). 
At higher coverage, the film enters a domain-wall phase (DW), with stripes of the $C_{1/3}$ solid separated by (superheavy) domain walls.\cite{Bruch97} 

At even higher coverage, we observe a change from striped to hexagonal network of (heavy) domain walls. The possibility of a transition between these two domain-wall types was already raised by Greywall.\cite{Greywall93} This network becomes denser with increasing coverage, ending with a commensurate solid ($C_{7/16}$) for $\rho_{7/16}=0.0835\pA$, where 7/16 of the adsorption sites are occupied. This structure is also found  in diffraction experiments of 
$D_2$ on graphite,\cite{Freimuth90} but had not yet been predicted for Helium. Greywall proposed a particular commensurate structure, around a coverage 0.820 $\pA$, which differs only by $1.8\%$ from  $\rho_{7/16}$. Thus, the signal observed in his heat capacity measurements may stem from the $C_{7/16}$ solid. For densities above $\rho_{7/16}$ we find an incommensurate solid, in agreement with experiments and previous calculations.\cite{Pierce00}
%
%%%%%%%%%%%%%%
%%%%%%%

\begin{figure}[t]
\center
  \includegraphics[width=0.48\textwidth]{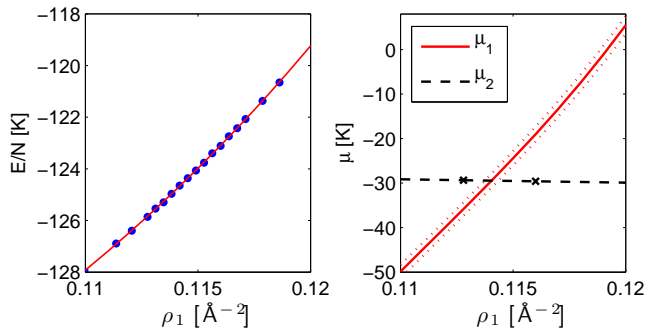}
  \caption{(Color online) Left plot: Energy per particle versus coverage of simulations with a constant number of particles (56) and varying volume. The red line is a polynomial fit of degree $4$. Right plot: The red curve corresponds to the chemical potential in the first layer, with an error bar given by the red dotted lines. The black dashed line indicates the chemical potential, at which the second layer becomes populated. The error bars are smaller than the symbol sizes. The intersection yields the layer promotion density $\rholp=0.1140 \pm 0.0003 \pA$.}
\label{layerprom}
\end{figure}

Next, we determine the layer promotion density $\rholp$, at which the second layer starts to become populated. 
At equilibrium, the chemical potentials of the first and the second layer are equal, i.e., 
$\mu_1(\rho_1, \rho_2) = \mu_2(\rho_1,\rho_2)$,
where $\rho_1$ and $\rho_2$ are the densities in the first and second layer respectively. 
Layer promotion occurs at specific value of the chemical potential $\mu^{lp}$ at which the second layer density $\rho_2$ jumps from zero to a finite value $\rho_2^{\text{min}}$. Thus, we can find $\rholp$ by solving the equation
\begin{equation}
\label{eq:rholp}
\mu^{lp}=\mu_1(\rholp, 0) = \mu_2(\rholp,\rho_2^{\text{min}}).
\end{equation}
In order to determine $\mu_1(\rho_1)$, we performed simulations with a fixed number of particles, and with the first layer initialized as a triangular solid. Density scans are obtained by varying the area of the simulation cell. It is
\begin{equation}
\mu_1(\rho_1) = e(\rho_1) + \rho_1 \frac{de}{d\rho_1}, 
\end{equation}
with $e=E/N$ the energy per particle, where we estimate the derivative
 from a polynomial fit of degree $4$ to $e(\rho)$ as shown in Fig. \ref{layerprom}. We determine the chemical potential of the second layer at the minimum coverage $\rho_2^{\text{min}}$ by taking the thermalized first layer simulations and switch to a grand canonical simulation, where we gradually increase $\mu$ until the second layer becomes populated. We find $\mu_2$ = $-$29.6 $\pm$ 0.3 K (essentially independent of $\rho_1$).
The solution of Eq.~\eqref{eq:rholp} for $\rholp$ is given by the intersection of the two chemical potential curves in the right panel in Fig. \ref{layerprom}. %, yielding $\rholp$=0.1140 $\pm$ 0.0003 $\pA$. %Comparing results from simulations with 30 and 56 particles we find that finite size effects are negligible.  
Comparing results from different simulations (30 and 56 particles), we find $\rholp$=0.1140 $\pm$ 0.0003 $\pA$. 
%
%%%%%%%%%%%%%%%%%%%%%%%%%%%%%%%%%%%%
% show_energy_a074.m
%%%%%%%%%%%%%%%%%%%%%%%%
%
%

This value is in agreement with previous simulation results of Whitlock,\cite{Whitlock98} who used an effective potential for second layer particles. It is also compatible with neutron diffraction experiments and heat capacity measurements,\cite{Bretz73,Polanco78, Lauter87, Carneiro81} where values in the range 0.112-0.115 $\pA$ were found. 

Greywall and Busch determined a value of 0.12 $\pA$ from heat capacity measurements, 
\cite{Greywall91} higher than our result. However, in Ref. \onlinecite{Greywall93} Greywall pointed  out that there is an ambiguity in his coverage scale by several percent. Thus, our layer promotion density is consistent with their value, taking their uncertainty into account. 
We have tested the sensitivity of our result upon deepening the attractive well  of \hef-graphite
potential  by $10\%$; we find  $\rholp$ = 0.1165 $\pm$ 0.0005 $\pA$ in this case, still lower 
than that of Greywall and Busch.  In fact, the value of density corresponding to first layer promotion proposed by Greywall and  Busch, is only observed by making the potential more attractive by over 
20\%, a correction which seems unlikely, as we discuss below.

\section{Results: Second Layer}

%%%%%%%%%%%
%Second layer results
%%%%%%%%%%%
%%%%%%%%%%%%%%%%%%%%%%%
\begin{figure}[t]
\center
     \includegraphics[width=0.48\textwidth]{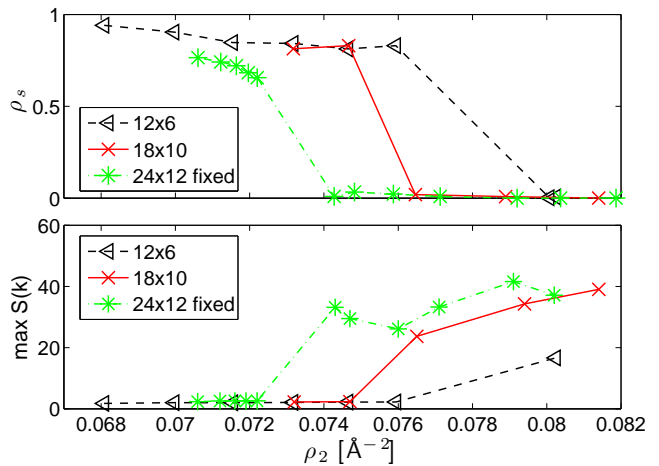}
  \caption{(Color online)  Upper panel: \hef\ superfluid density. Lower panel: peak values of the static structure factor. Data shown are for the second layer a temperature $T=0.5$ K. The system sizes are given as multiples of the first layer unit cell. The first layer density is $\rho_1=0.1202\ \pA$ (0.1270 $\pA$) in simulations with active (fixed, stars) first layer particles.  Statistical errors are of the order of symbol sizes. }
\label{fig:rhos_sk}
\end{figure}
%%%%%%%%%%%%%%%%%%%%%%%%

We now discuss the physics of the second adsorbed helium layer. Adsorption of successive layers causes a compression of the first adlayer, experimentally estimated between four \cite{Lauter87} and six percent.\cite{Greywall91} In this work, we performed simulations for the second layer based on different first layer densities, ranging from 0.1164  to 0.1270 $\pA$, the latter being the value proposed by Greywall and Busch,\cite{Greywall91, Greywall93} and assumed by Pierce and Manousakis in their PIMC simulations.\cite{Pierce98,notecell} 

On varying the density of the first layer in the above range, the physics of the second layer does not change qualitatively;  typical results are shown in Fig. \ref{fig:rhos_sk}. Specifically, we {\it only} find a superfluid and an incommensurate solid phase, separated by a first-order phase transition. We do {\it not} observe a commensurate solid phase sandwiched between the liquid and the incommensurate crystal, in contrast to the prediction of Greywall and Busch.

The 2d equilibrium density of the liquid second layer, in the $T\to 0$ limit,  is estimated at 0.046(3) $\pA$, essentially independent of first layer density.  In order to give an idea of the weakness of the adsorption potential seen by second layer atoms, one may note that the above equilibrium density is indistinguishable from that of  purely two-dimensional \hef\,,\cite{Gordillo98}  and significantly lower than that of a \hef\  {\it mono}-layer on a lithium substrate (the weakest known substrate wetted by \hef).\cite{szybisz} This liquid film turns superfluid at low $T$, as we established by direct computation of the superfluid density $\rho_S$, based on the well-known winding number estimator.\cite{Ceperley95} 

Assuming a first layer density of 0.1202 $\pA$ (corresponding to a 5\% compression), the onset value of coverage for the occurrence of superfluidity is therefore 0.166(3) $\pA$ (Fig.  \ref{fig:firstlayer}). As shown in Fig. \ref{fig:rhos_sk} (crosses and triangles), the superfluid density vanishes at a 2d density of 0.076 $\pm$ 0.002 $\pA$, corresponding to a 
coverage of 0.196 $\pm$ 0.002 $\pA$, at which the incommensurate crystal phase appears. These coverages are altogether  consistent with existing measurements, once experimental uncertainties are properly taken into account.\cite{Crowell}

In the heat-capacity measurements of Greywall and Busch, a peak appears around a coverage of  0.197 $\pA$, which they associate with a commensurate structure.
%, because the peak shows little density dependence and exists only over a very restricted 
% range of second-layer densities. 
They assumed a compressed first layer density of 0.1270 $\pA$. From the ratio of first and second layer densities ($\approx 4/7$), Greywall conjectured a $\sqrt{7}\times\sqrt{7}$ registered structure with one-quarter of the second layer atoms located directly above the first layer atoms; this has also been proposed for the second layer of \het on graphite.\cite{Elser89} 
To our knowledge, however, no direct experimental evidence for the commensurate phase has been reported. One possible scenario is that 
the observed peak in the heat capacity is a signature of the incommensurate phase, which in our simulations appears at the same coverage. 
 %
%If we start the simulation with the particles initialized in the commensurate structure, we observe %melting into the superfluid phase. 
 
%%%%%%%%%%%%%%%%%%%%%%%
% 
\begin{figure}[t]
\center
  \includegraphics[width=0.48\textwidth]{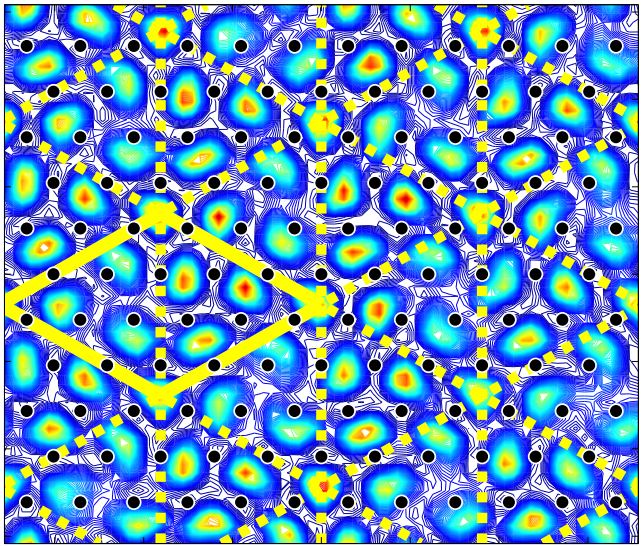}
  \caption{(Color online) Snapshot of the commensurate 7/12 solid in the second layer, which only appears in simulations with fixed first layer particles with a first layer density of $\rho_1=0.1270\pA$. The black dots mark the positions of the first layer particles (repulsion sites). Second layer particles found at the intersections of the thick dotted (yellow) lines are located above the middle of three neighboring first layer particles. The (yellow) thick line marks a unit cell of the 7/12 solid. The \hef\ superfluid density in this phase is zero. }
\label{snapshot712}
\end{figure}
%%%%%%%%%%%%%%%%%%%%%%%%

In the PIMC simulations of Ref. \onlinecite{Pierce98}, the 4/7  commensurate structure was observed (with a different positioning of the second layer atoms with respect to the first layer), assuming the first layer density proposed by Greywall and Busch. However, as mentioned above, in these simulations first layer particles were held {\it fixed}, i.e. their zero-point motion was neglected. 
On performing the same simulation, we also observe a commensurate structure (Fig. \ref{fig:rhos_sk}), albeit at a slightly different filling (7/12). It consists of a triangular lattice rotated by an angle of $10.89^\circ$ with respect to the first layer, as shown in the snapshot in Fig. \ref{snapshot712}. The difference in density compared to the $4/7$ filling is only $\approx 2\%$. The superfluid density of such a commensurate phase is {\it zero}, i.e., no evidence of a possible supersolid phase is found (see Fig. \ref{fig:rhos_sk}).

\begin{figure}[t]
\center
\includegraphics[width=0.48\textwidth]{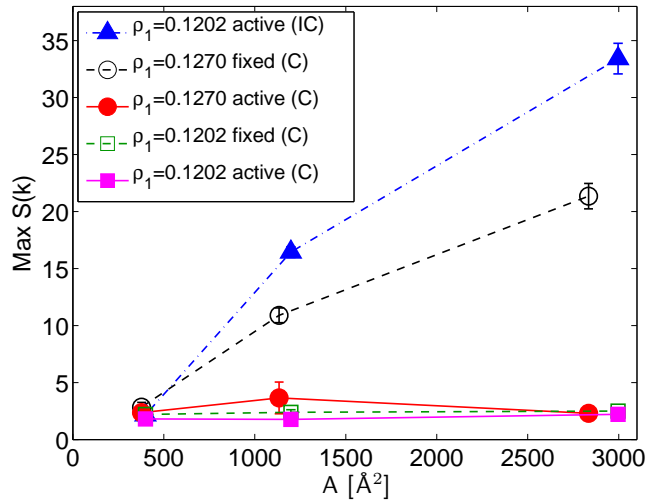}
    \caption{(Color online) Scaling of the structure factor peak with system size A (area) at temperature $T=0.5\K$. Results obtained from active/fixed first layer particles for two different first layer densities $\rho_1$ are shown. The incommensurate solid (triangles, IC) is always stable, independent of the first layer density. The second layer density is $\rho_2=0.080\pA$ in this example.
  For all the other data points the coverage corresponds to the one of the commensurate $7/12$ solid. 
  Only in the case of fixed first layer particles and highest first layer density $\rho_1=0.1270 \pA$ the commensurate solid is stable in the thermodynamic limit. }
\label{fig:sfact}
\end{figure}

\begin{figure}[t]
\center
\includegraphics[width=0.40\textwidth]{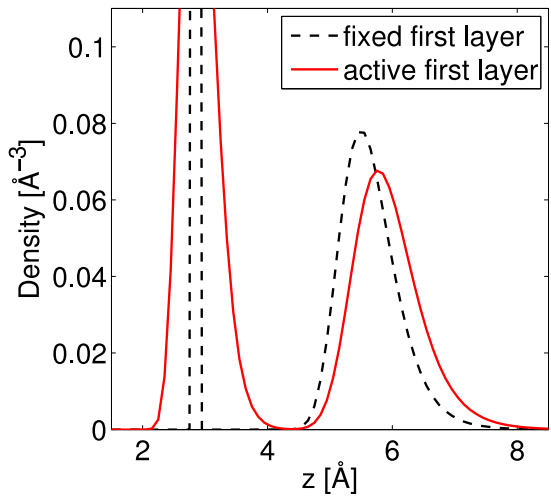}
    \caption{(Color online) Comparison between the density profiles of a system with fixed (dashed line) and active (full line) first layer particles at a total coverage $\rho=0.2117~\pA$. % In the former case the zero-motion of the first layer particles is neglected. As a consequence the second layer moves closer to the substrate
    In the latter case the second layer is displaced by $\approx$ 0.3 $\textrm{\AA}$  away from the substrate due to the zero-point motion of the first layer particles. 
    }
\label{fig:densz_compare}
\end{figure}

As shown in Figs. \ref{fig:rhos_sk} and \ref{fig:sfact}, no commensurate phase arises if first layer particles are simulated {\it explicitly}, even if the (relatively high) first layer density of 0.127 $\pA$ utilized in Ref. \onlinecite{Pierce98} is assumed.
This is because, due to zero-point motion, first layer atoms occupy a larger region of space than predicted classically. Thus, second layer atoms are slightly pushed away from the first layer (see Fig. \ref{fig:densz_compare}), which reduces the substrate attraction by $\sim$ 3 K,  enough to de-stabilize the commensurate phase (see also Ref. \onlinecite{Whitlock99}). It should also be noted that, even with fixed first layer particles, no commensurate phase is observed if a first layer density lower than 0.127 $\pA$ is assumed (Fig. \ref{fig:sfact}).

%Therefore, the approximation of fixed particles in the layer is not appropriate in this case. 

%%%%%%%%%%%%%%%%%%%%%%%
% 

Regardless of whether first layer particles are held fixed or not, at sufficiently high coverage an {\it incommensurate} crystalline phase forms (see Fig. \ref{fig:rhos_sk}). Such a phase is also not superfluid, consistently with what is now regarded as a fairly general theoretical statement.\cite{Prokofev05}

Just like for the issue of first layer promotion, we have explored the possibility that a revision of the helium-graphite potential might indeed stabilize the commensurate second layer phase which we do not observe using the Carlos-Cole potential as originally proposed in Ref. \onlinecite {Carlos79}. Simulations with a $10\%$ deeper substrate
potential also failed to yield a stable commensurate solid, except at the very largest  
first layer density $\rho_1$ = 0.1270 $\pA$ (which is however not compatible with our value for 
$\rholp$), where the 7/12 commensurate solid becomes stable. We discuss below whether a quantitative revision of the potential seems justified, in light of known experimental facts. 

%%%%%%%%%%%%%%%%%%%%%%%%%%%%%

%%%%%%%%%%%%%%%%%%%%%%%
% 
%%%%%%%%%%%%%%%%%%%%%%%%%%%%%

%
\section{Conclusions}
We have carried out a thorough computational study of the first and second layers of \hef\ adsorbed on a graphite substrate, based on the most realistic interaction potentials currently available, utilized in all previous simulation work. For the first layer, our results largely confirm those of others, and are consistent with experimental results.   

%using the best available interaction potentials 
Conversely, our simulations show {\it no} commensurate second layer phase, in contrast to the interpretation of heat capacity measurements by Greywall and Busch \cite{Greywall91, Greywall93}. Rather, we only observed a liquid layer (superfluid at low temperature), which crystallizes at high density to form an incommensurate solid. The physical behavior of the second layer is thus  very close to that of a purely two-dimensional system, also looking at the narrow spread of the \hef\ density in the perpendicular direction and at the non-existent overlap of first and second layer particles shown in Fig. 
\ref{fig:densz_compare}, pointing to the absence of interlayer quantum exchanges. It is also worth pointing out the similarity between the physics of the  second layer of \hef\  on graphite and that of a mono-layer adsorbed on a lithium substrate.\cite{szybisz,toigo}

Our prediction of  no commensurate phase,  is at variance with previous numerical work by Pierce and Manousakis, who observed it instead in their PIMC simulations. It need be stressed, however, that there is no significant {\it numerical} disagreement between our results and those of Ref. \onlinecite{Pierce98}. Our different conclusion directly stems from the fact that, unlike Pierce and Manousakis, we did {\it not} keep first layer particles fixed, but rather simulated their zero-point motion explicitly. This fact alone accounts for (most of) the difference between the physical outcomes of the two studies. 
%Simulations in which first layer particles are held fixed yielded the same result as in Ref. \onlinecite{Pierce98} (using the same first layer density utilized by Pierce  and Manousakis). 
It is worth noting that the possible importance of the role of quantum fluctuations of inner layer particles had already been suggested by other authors.\cite{Whitlock99}

The absence of a commensurate phase suggests that either the experimental data have to be reinterpreted, or the microscopic model adopted so far, chiefly the helium-graphite potential, might have to  undergo significant revisions. We have studied this scenario to same length, in this work, notably by rendering the attractive well of the potential deeper. If the attraction is increased by some 10\% (while fully retaining in the simulation the zero-point motion of first layer \hef\ atoms) a commensurate crystalline phase is observed  {\it only} if a density of 0.127 $\pA$ is assumed for the first layer, i.e., the value proposed by Greywall and Busch. This value is consistent with the first layer promotion density computed in this work with such a revised potential (i.e., 10\% more attractive), {\it only} on assuming a compression of the inner adlayer of some 9\% upon adsorption of successive layers. This degree of compression is substantially above that estimated by most experimental studies.

In any case, a revision of such a quantitative degree of the helium-graphite potential proposed by Carlos and Cole seems doubtful,  in view of the good agreement with experimental data presented in 
Ref. \onlinecite{Carlos79}. 

% by more than $10\%$. To 
%I 
Finally, no finite superfluid signal is seen in any of the crystalline phases observed, either the incommensurate or  the commensurate (the latter, as explained above, merely occurs as the result of treating inner layer particles as fixed in the simulation). Thus, we conclude that  this system is no realistic candidate for the observation of a supersolid phase.

\section*{ACKNOWLEDGMENTS}
We acknowledge inspiring discussions with N. Prokof'ev, J. Ny\'eki and J. Saunders. Simulations were performed on the Brutus Beowulf cluster at ETH Zurich. This work was supported by the Natural Science and Engineering Research Council of Canada under grant G12120893, and the Swiss National Science Foundation.

%%%%%%%%%%%%%%%%%%%%%%%%%%%%%%%%%%%%%%%%%%%%%%%%%%%%%

%\bibliographystyle{../diss/chapters/gianni_prl}
%\bibliography{../diss/chapters/helium}

\begin{thebibliography}{25}
\bibitem[{\citenamefont{Saunders}()}]{Saunders}
\bibinfo{author}{\bibfnamefont{J.}~\bibnamefont{Saunders}},
  \bibinfo{note}{private communication. It should be noted that there is a qualitative difference between a system in which crystalline order arises from the spontaneous breaking of translational symmetry, and one in which it is imposed by an external potential, as would be the case for an adsorbed film, registered with the underlying substrate.}
\bibitem[{\citenamefont{Piercenote}(1998)}]{Piercenote}
\bibinfo{note}{\bibfnamefont{In Ref. \onlinecite{Pierce98}  a slightly different version of the Aziz potential was used, which does not deviate significantly from the one used here.}}
\end{thebibliography}

\end{document}